\documentclass[journal]{IEEEtran} 
\ifCLASSINFOpdf
\else
\fi

\usepackage{moreverb}
\usepackage{graphicx}
\usepackage{amsmath}
\usepackage{amsthm}
\usepackage{amssymb}
\usepackage{epsfig}
\usepackage{epstopdf}
\usepackage{float}
\usepackage{algorithm}
\usepackage{algpseudocode}
\usepackage{blkarray}
\usepackage{wrapfig}
\usepackage{cite}
\usepackage{color}
\usepackage{multirow}
\usepackage[subpreambles=true]{standalone}
\usepackage{mathptmx}
\usepackage[latin1]{inputenc}
\usepackage[T1]{fontenc}
\usepackage{nccmath}

\begin{document}
\title{\LARGE Device-to-Device Communication Facilitating Full-Duplex Cooperative Relaying Using Non-Orthogonal Multiple Access}
\author{Mohammed~Belal~Uddin, Md.~Fazlul~Kader,~\IEEEmembership{Senior Member,~IEEE} and Soo~Young~Shin,~\IEEEmembership{Senior Member,~IEEE} \vspace{-4ex}
	\thanks{Manuscript received XXX, XX, 2018; revised XXX, XX, 2018. This work was supported by Priority Research Centers Program through the National Research Foundation of Korea(NRF) funded by the Ministry of Education, Science and Technology (2018R1A6A1A03024003).}
	\thanks{Mohammed Belal Uddin  and Soo Young Shin are with the WENS Laboratory, Department of IT Convergence Engineering, Kumoh National Institute of Technology, Gumi 39177, South Korea
		(email: ahad.belal@kumoh.ac.kr, wdragon@kumoh.ac.kr).}%
	\thanks{Md. Fazlul Kader is with the Department of Electrical and Electronic Engineering, University of Chittagong, Chittagong-4331, Bangladesh 
		(email: f.kader@cu.ac.bd).}
	
}
\markboth{Submitted Journal,~Vol.~xx, No.~xx, xx~2018}%
{}
\maketitle
\begin{abstract}
This letter presents a device-to-device (D2D) enabling cellular full-duplex (FD) cooperative protocol using non-orthogonal multiple access (NOMA), where an FD relay assists in relaying NOMA far user's signal and transmits a D2D receiver's signal simultaneously. The ergodic capacity, outage probability, and diversity order of the proposed protocol are theoretically investigated under the realistic assumption of imperfect self and known interference cancellation. The Outcome of the investigation demonstrates the performance gain of the suggested protocol over conventional FD cooperative NOMA system.

\end{abstract}
\begin{IEEEkeywords}
Non-orthogonal multiple access, ergodic sum capacity, full-duplex, device-to-device communication.
\end{IEEEkeywords}
\IEEEpeerreviewmaketitle
\section{Introduction}
\IEEEPARstart{N}{on-orthogonal} multiple access (NOMA) enables a transmitter to transmit multiple signals concurrently to multiple receivers having distinguished channel quality. NOMA is also combinable with other technical features like cooperative communication, device-to-device (D2D) communication, half/full duplex relaying, etc. Hence, it is considered as one of the promising radio access technologies for future wireless communication~\cite{1, 2}. NOMA adopted full-duplex relay (FDR) aided cooperative communication (FDCC) can efficiently enhance spectral efficiency, ergodic capacity (EC), and signal reliability~\cite{3}. An FDR either can be a dedicated relay~\cite{3} or a relay-like user~\cite{4}. In order to increase the outage performance of a far user, a simple FDCC protocol was considered~\cite{4} wherein a NOMA near user relayed the signal to a far user. In~\cite{5}, a non-cooperative NOMA protocol consisting of a pair of cellular users and a pair of D2D users in a cell was devised and impact of interference from a D2D transmitter on cellular user's performance was investigated. 
An FDCC system was analyzed in~\cite{6}, where a dedicated FDR was used to transmit signal to a far user. Outage probability (OP) and ergodic sum capacity (ESC) were shown higher than the existing corresponding half-duplex cooperative relaying system. Focusing on achieving better spectral efficiency than~\cite{6}, an integrated D2D and cellular communication protocol is proposed in this letter. A succinct description of the contribution of this work is summarized as: (i) A D2D facilitating and an FDR assisted cooperative NOMA (termed as DFC-NOMA) protocol is proposed, where an FDR relays the far user's signal and acts as a D2D transmitter as well. (ii) A closed-form and an asymptotic (Asm.) expressions of each user's EC, OP, and system's ESC are analyzed over Rayleigh fading channel. The diversity order (DO) of each user is also investigated. (iii) Finally,  performance gain of the proposed DFC-NOMA over conventional full-duplex cooperative NOMA (FC-NOMA)~\cite{6} is shown by theoretical analysis and justified by simulation.
\begin{figure}[t]
	\centering
	\includegraphics[width=3in,height=2in,keepaspectratio]{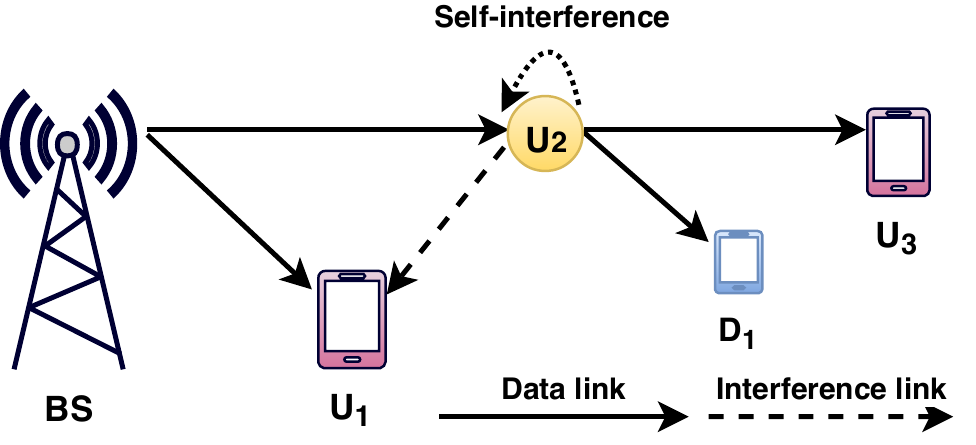}
	\caption{Proposed network model.}
	\label{Fig1}
\end{figure}
\section{System Model and Protocol Description}
As shown in Fig.~\ref{Fig1}, a downlink DFC-NOMA is considered that consists of a base station (BS), two cellular users U$_1$ (near user) and U$_3$ (far user), one full-duplex relay acting user (U$_2$) for conveying signal to U$_3$ from BS, and a D2D user (D$_1$). Due to the blockage and hindrance to signal propagation, BS-to-U$_3$ link is considered unavailable~\cite{6} and the remaining links of the system are available. In~\cite{6}, the relay retransmits the decoded symbol only to far user. Contrarily, in the proposed DFC-NOMA, U$_2$ (acting as a relay) retransmits the decoded symbol to U$_3$ and its own symbol to D$_1$ at the same time, by exploiting NOMA. 
The channel coefficient experiences Rayleigh fading between two nodes \textit{i} and \textit{j} will be complex Gaussian random variable with zero mean and variance $\lambda_{ij}$ that can be denoted by $h_{ij}\sim \mathcal{C}\mathcal{N} (0, \lambda_{ij})$; \textit{i}, \textit{j} $\in$ \{BS, U$_1$, U$_2$, U$_3$, D$_1$\}, and \textit{i} $\neq$ \textit{j}. 
Hence, the channel gain  $g_{ij}$ = $|h_{ij}|^2$ will be the exponentially distributed random variable with scale parameter of $\lambda_{i ,j}$ \cite{3}. Subscripts $b$, 1, 2, 3, and $d$ are respectively used for BS, U$_1$, U$_2$, U$_3$, and D$_1$. Each node excluding U$_2$ is equipped with single antenna. As U$_2$ operates in full-duplex mode, it is equipped with one transmit and one receive antennas. So, due to the simultaneous transmission and reception of signals, the relay suffers from a self-interference (SI) that can be subsided by applying multi-stage SI cancellation technique~\cite{7}. In practice, SI can not be removed perfectly. Therefore, imperfect SI cancellation is considered that causes residual SI at relay which is symbolized by $\tilde{h}_{22}\sim \mathcal{C}\mathcal{N} (0, \varsigma_1\lambda_{22})$ with zero mean and $\varsigma_1\lambda_{22}$ variance. Hence, the channel gain of residual SI at U$_2$ is $\tilde{g}_{22}=|\tilde{h}_{22}|^2$. The variable $\varsigma_1$ limiting as $0 \leq \varsigma_1 \leq 1$ specifies the effect of remaining SI. Devices are assumed to be located in such a way that the channel gains can be considered as $g_{b2} \ll g_{b1}$ and $g_{23} \ll g_{2d}$.
At any instant of time ($\textit{t}$-th slot), the BS transmits a composite downlink NOMA signal $s_c$[$\textit{t}$] = $\sqrt{\theta_1p_b}s_{1}[\textit{t}]+\sqrt{\theta_3p_b}s_{3}[\textit{t}]$where s$_1$ information is intended to reach U$_1$ directly and s$_3$ information is intended to be reached to U$_3$ via the help of U$_2$. Parameters $p_b$, $\theta_1$, and $\theta_3$ refer to the total transmit power of BS, associated power allocation factor with s$_1$, and associated power allocation factor with s$_3$, respectively, where $\theta_1$ + $\theta_3$ = 1 and $\theta_1$ < $\theta_3$ is maintained due to the channel gain difference between BS-to-U$_1$ and BS-to-U$_2$ as stated earlier. The relay U$_2$ decodes $s_3$ at the direct phase and retransmits this decoded symbol along with it's own signal $s_2$ with a processing delay $\nu\ge1$ at cooperative phase. According to the property of downlink NOMA, this transmit signal can be written as $s_r$[$\textit{t}-\nu$] = $\sqrt{\theta_2p_u}s_{2}[\textit{t}-\nu]+\sqrt{\theta_3^\prime p_u}s_{3}[\textit{t}-\nu]$, where $\theta_2$ + $\theta_3^\prime$ = 1, $\theta_2$ < $\theta_3^\prime$, $p_u$ is the total transmit power of U$_2$, $\theta_2$ and $\theta_3^\prime$ are the allocated power with $s_{2}$ and $s_{3}$ informations, respectively. 
After receiving the transmitted signal from BS, relay decodes the information $s_{3}$ by treating $s_{1}$ as noise and the related signal-to-interference-plus-noise ratio (SINR) at U$_2$ is given by
\begin{align}
\gamma_{b2}^{s_3} =& \frac{\theta_3\rho_b g_{b2}}{\theta_1\rho_b g_{b2}+\rho_u\tilde{g}_{22}+1}, \label{eq1}
\end{align}
where $\rho_b=\frac{p_b}{\sigma_{2}^2}$ and $\rho_u=\frac{p_u}{\sigma_{2}^2}$ respectively represent the transmit SNR of BS and U$_2$; $\sigma_{2}^2$ is the variance of additive white Gaussian noise (AWGN) at U$_2$. Let, AWGN at all receivers are identical. So, the variance of AWGN will be equal for all receivers. When the near user U$_1$ receives the transmitted signal from BS, it also gets the composite signal from U$_2$ as interference. During the self-signal decoding U$_1$ decodes U$_3$'s signal $s_3$ first and then performs successive interference cancellation (SIC) to obtain it's own signal $s_1$. So, the information $s_3$ in interference signal from U$_2$ is known to U$_1$ and can remove it by applying known interference cancellation technique. But U$_1$ can not remove $s_2$ from interference signal as this is unknown to U$_1$. Therefore, by considering imperfect cancellation of known interference $s_3$ and no cancellation of unknown interference $s_2$, the channel coefficient of interference link from U$_2$ to U$_1$ can be modeled as $\tilde{h}_{21}\sim \mathcal{C}\mathcal{N} (0, \varsigma_2\theta_2\lambda_{21}+\varsigma_3\theta_3^\prime\lambda_{21})$, where the parameters $\varsigma_2$ (=1) and $\varsigma_3$ ($0 \leq \varsigma_3 \leq 1$) refer level of residual interference.
Accordingly, the received SINRs related to informations $s_3$ and $s_1$ at U$_1$ are respectively obtained as 
\begin{align}
\gamma_{b1}^{s_{3}} =& \frac{\theta_3\rho_b g_{b1}}{\theta_1\rho_b g_{b1}+\rho_u\tilde{g}_{21}+1}, \label{eq2}\\
\gamma_{b1}^{s_1} =& \frac{\theta_1\rho_b g_{b1}}{\rho_u\tilde{g}_{21}+1}, \label{eq3}
\end{align} 
where $\tilde{g}_{21}=|\tilde{h}_{21}|^2$. 
At the cooperative phase, the far/weak user U$_3$ receives the signal transmitted from U$_2$ and decodes it's information $s_3$ by treating $s_2$ as noise. So, the related SINR for $s_3$ at U$_3$ is given by  
\begin{align}
\gamma_{23}^{s_3} =& \frac{\theta_3^\prime\rho_u g_{23}}{\theta_2\rho_u g_{23}+1}. \label{eq4}
\end{align}
D2D user D$_1$ also receives the transmitted signal from U$_2$ at cooperative phase. First, D$_1$ needs to decode $s_3$ by treating $s_2$ as noise. Then after performing SIC process, it decodes it's own information $s_2$. Hence, SINRs related to $s_3$ and $s_2$ at D$_1$ can be respectively expressed as
\begin{align}
\gamma_{2d}^{s_3} =& \frac{\theta_3^\prime\rho_u g_{2d}}{\theta_2\rho_u g_{2d}+1}, \label{eq5}\\
\gamma_{2d}^{s_2} =& \theta_2\rho_u g_{2d}. \label{eq6}
\end{align}
Using (\ref{eq3}) and (\ref{eq6}), the achievable rate of U$_1$ and D$_1$ are respectively written by  
\begin{align}
C_{1} =& \text{log}_2\left(1+\gamma_{b1}^{s_1}\right), \label{eq7}\\
C_{d} =& \text{log}_2\left(1+\gamma_{2d}^{s_2}\right). \label{eq8}
\end{align} 
Moreover, the achievable rate of U$_3$ can be obtained by using (\ref{eq1}), (\ref{eq2}), (\ref{eq4}), and (\ref{eq5}) as
\begin{align}
C_{3} =& \text{log}_2\left(1+\text{min}\{\gamma_{b2}^{s_3}, \gamma_{b1}^{s_3}, \gamma_{23}^{s_3}, \gamma_{2d}^{s_3}\}\right). \label{eq9}
\end{align}  
Finally, the overall achievable capacity of the proposed DFC-NOMA can be calculated as 
\begin{align}
C_{\text{cap.}}^{\text{pro.}} =& C_{1}+C_{d}+C_{3}. \label{eq10}
\end{align} 
\section{Performance Analysis}
\subsection{Capacity Analysis}
\subsubsection{Ergodic Capacity of U$_1$}
Let $Q\triangleq\frac{\theta_1\rho_b g_{b1}}{\rho_u\tilde{g}_{21}+1}$. Using the definition of cumulative distribution function (CDF), $F_{Q}(q)=\text{Pr}\{\frac{\theta_1\rho_b g_{b1}}{\rho_u\tilde{g}_{21}+1}<q\}$, the CDF of Q can be derived as 
\begin{align}
F_{Q}(q) =& 1-e^{-\frac{q}{\theta_1\rho_b\lambda_{b1}}}\left(1+\frac{\rho_u\left(\varsigma_2\theta_2+\varsigma_3\theta_3^\prime\right)\lambda_{21}q}{\theta_1\rho_b\lambda_{b1}}\right)^{-1}. \label{eq11}
\end{align}  
Considering $\alpha=\frac{1}{\theta_1\rho_b\lambda_{b1}}$, $\beta=\rho_u\left(\varsigma_2\theta_2+\varsigma_3\theta_3^\prime\right)\lambda_{21}$, and applying (\ref{eq11}) to (\ref{eq7}), the closed-form solution of U$_1$'s EC can be calculated as  
\begin{align}
\bar{C}_{1} =& \frac{1}{\text{ln2}}\displaystyle \int^{\infty}_{0}\frac{1-F_{Q}(q)}{1+q}\,dq = \frac{1}{\text{ln2}}\displaystyle \int^{\infty}_{0}\frac{1}{\left(1+q\right)\left(1+\alpha\beta q\right)}e^{-\alpha q}\,dq \nonumber\\
=& \frac{\text{log}_2e}{1-\alpha\beta}\left[e^{\frac{1}{\beta}}\text{Ei}\left(-\frac{1}{\beta}\right)-e^\alpha\text{Ei}\left(-\alpha\right)\right], \label{eq12} 
\end{align} 
where $\text{Ei}\left(y\right)=\displaystyle \int^{y}_{-\infty}\frac{e^a}{a}\,da$ expresses the exponential integral function~\cite{8}. In case of $\rho_b\rightarrow\infty$ and $\rho_u\rightarrow\infty$, $\text{Ei}\left(-x\right)\approx E_c+\ln\left(x\right)$~\cite{8} and $e^x\approx1+x$ can be applied to derive the asymptotic EC of U$_1$ as follows.
\begin{align}
\bar{C}_{1}^\infty =& \frac{\text{log}_2e}{1-\alpha\beta}\left[\left(1+\frac{1}{\beta}\right)\left\{E_c-\ln\left(\beta\right)\right\}-\left(1+\alpha\right)\left\{E_c+\ln\left(\alpha\right)\right\}\right], \label{eq13} 
\end{align}
where $E_c$ represents the Euler constant. 
\subsubsection{Ergodic Capacity of D$_1$}
Assuming $T\triangleq\gamma_{2d}^{s_2}=\theta_2\rho_u g_{2d}$, the CDF $F_{T}(t)$ can be obtained as  
\begin{align}
F_{T}(t) =& 1-e^{-\frac{t}{A_d}}. \label{eq14}
\end{align}
where $A_d=\theta_2\rho_u\lambda_{2d}$. Using (\ref{eq8}) and (\ref{eq14}), the EC of D$_1$ is written as    
\begin{align}
\bar{C}_{d} =& -\frac{1}{\ln2}\displaystyle \int^{\infty}_{0}\frac{1}{\left(1+t\right)}e^{-\frac{t}{A_d}}\,dt = -\frac{1}{\ln2}e^{\frac{1}{A_d}}\text{Ei}\left(-\frac{1}{A_d}\right). \label{eq15} 
\end{align}
By following similar way as applied for $\bar{C}_{1}^\infty$, the asymptotic EC of D$_1$ also can be obtained as follows. 
\begin{align}
\bar{C}_{d}^\infty =& \log_2\left(\frac{1}{e}\right)\left(1+\frac{1}{A_d}\right)\left\{E_c+\ln\left(\frac{1}{A_d}\right)\right\}. \label{eq16} 
\end{align}
\subsubsection{Ergodic Capacity of U$_3$}
Let $X\triangleq\frac{\theta_3\rho_b g_{b2}}{\theta_1\rho_b g_{b2}+\rho_u\tilde{g}_{22}+1}$, $Y\triangleq \frac{\theta_3\rho_b g_{b1}}{\theta_1\rho_b g_{b1}+\rho_u\tilde{g}_{21}+1}$, $Z\triangleq\frac{\theta_3^\prime\rho_u g_{23}}{\theta_2\rho_u g_{23}+1}$, $W\triangleq\frac{\theta_3^\prime\rho_u g_{2d}}{\theta_2\rho_u g_{2d}+1}$, and $R\triangleq\text{min}\left(X,Y,Z,W\right)$. The CDF of \textit{X}, \textit{Y}, \textit{Z}, and \textit{W} are respectively calculated as 
\begin{align}
F_{X}(x) =& 1-e^{-\frac{x}{\left(\theta_3-\theta_1x\right)\rho_b\lambda_{b2}}}\left(1+\frac{\varsigma_1\rho_u\lambda_{22}x}{\left(\theta_3-\theta_1x\right)\rho_b\lambda_{b2}}\right)^{-1}, \label{eq17}\\
F_{Y}(y) =& 1-e^{-\frac{y}{\left(\theta_3-\theta_1y\right)\rho_b\lambda_{b1}}}\left(1+\frac{\rho_u\left(\varsigma_2\theta_2+\varsigma_3\theta_3^\prime\right)\lambda_{21}y}{\left(\theta_3-\theta_1y\right)\rho_b\lambda_{b1}}\right)^{-1}, \label{eq18}\\
F_{Z}(z) =& 1-e^{-\frac{z}{\left(\theta_3^\prime-\theta_2z\right)\rho_u\lambda_{23}}}, \label{eq19}\\
F_{W}(w) =& 1-e^{-\frac{w}{\left(\theta_3^\prime-\theta_2w\right)\rho_u\lambda_{2d}}}. \label{eq20}
\end{align} 
Using (\ref{eq17}), (\ref{eq18}), (\ref{eq19}), and (\ref{eq20}), the CDF of $R$ is written as 
\begin{align}
F_{R}(r) =& 1-e^{-\frac{Dr}{\left(\theta_3-\theta_1r\right)}-\frac{Er}{\left(\theta_3^\prime-\theta_2r\right)}}\left[\frac{\left(\theta_3-\theta_1r\right)^2\rho_b^2\lambda_{b1}\lambda_{b2}}{\{A+Gr\}\{B+\left(\beta-J\right)r\}}\right], \label{eq21}
\end{align}
where $A=\theta_3\rho_b\lambda_{b2}$, $B=\theta_3\rho_b\lambda_{b1}$, $D=\frac{1}{\rho_b}\left(\frac{1}{\lambda_{b2}}+\frac{1}{\lambda_{b1}}\right)$, $E=\frac{1}{\rho_u}\left(\frac{1}{\lambda_{23}}+\frac{1}{\lambda_{2d}}\right)$, $G=\varsigma_1\rho_u\lambda_{22}-\theta_1\rho_b\lambda_{b2}$, and $J=\theta_1\rho_b\lambda_{b1}$. Putting in $\textstyle \int^{\infty}_{0}\log_2\left(1+r\right)f_{R}(r)\,dr=\frac{1}{\ln2}\textstyle \int^{\infty}_{0}\frac{1-F_{R}(r)}{1+r}\,dr$, the EC of U$_3$ can be represented as 
\begin{align}
\bar{C}_{3} =& \frac{1}{\text{ln2}}\displaystyle \int^{\infty}_{0}\frac{\left(\theta_3-\theta_1r\right)^2\rho_b^2\lambda_{b1}\lambda_{b2}e^{-\frac{Dr}{\left(\theta_3-\theta_1r\right)}-\frac{Er}{\left(\theta_3^\prime-\theta_2r\right)}}}{\left(1+r\right)\{A+\left(G-H\right)r\}\{B+\left(\beta-J\right)r\}}\,dr. \label{eq22} 
\end{align}
However, the closed-form solution of (\ref{eq22}) is not tractable~\cite{6}. Rather, it can be evaluated through the numerical integration. To find out the asymptotic solution, consider $\rho_b=\rho_u=\rho$, $\rho\rightarrow\infty$, $\theta_1=\theta_2$, $\theta_3=\theta_3^\prime$, and $R\triangleq\text{min}\left(X,Y,Z,W\right)\approx\text{min}\left(\frac{\theta_3}{\theta_1},\frac{\theta_3g_{b1}}{\theta_1g_{b1}+\tilde{g}_{21}}\right)\approx\text{min}\left(\frac{\theta_3}{\theta_1},N\right)$ under perfect SI cancellation. Using $n=\frac{g_{b1}\left(\theta_3-\theta_1n\right)}{\tilde{g}_{21}}$, the CDF of $N$  can be written as
\begin{align}
F_{N}(n) =& \frac{n\beta}{\rho_u\lambda_{b1}\left(\theta_3-\theta_1n\right)+n\beta}. \label{eq23}
\end{align} 
Using (\ref{eq23}) and following \cite{3}, the asymptotic EC of U$_3$ can be derived as   
\begin{align}
\bar{C}_{3}^\infty =& \displaystyle \int^{\infty}_{\frac{\theta_3}{\theta_1}}\log_2\left(1+\frac{\theta_3}{\theta_1}\right)f_{N}(n)\,dn+\displaystyle \int^{\frac{\theta_3}{\theta_1}}_{0}\log_2\left(1+n\right)f_{N}(n)\,dn \nonumber\\
=& \frac{1}{\text{ln2}}\displaystyle \int^{\frac{\theta_3}{\theta_1}}_{0}\frac{1-F_{N}(n)}{1+n}\,dn = \frac{1}{\text{ln2}}\displaystyle \int^{\frac{\theta_3}{\theta_1}}_{0}\frac{\chi-\psi n}{\left(\chi+n\right)\left(1+n\right)
}\,dn \nonumber\\
=& \frac{\log_2e}{1-\psi}\left\{\psi\left(1+\chi\right)\ln\frac{\left(\frac{\theta_3}{\theta_1}+\psi\right)}{\psi}-\left(\psi+\chi\right)\ln\left(\frac{\theta_3}{\theta_1}+1\right)\right\}, \label{eq24}
\end{align} 
where $\psi=\frac{\theta_3\rho_u\lambda_{b1}}{\beta-\theta_1\rho_u\lambda_{b1}}$ and $\chi=\frac{\theta_1\rho_u\lambda_{b1}}{\beta-\theta_1\rho_u\lambda_{b1}}$. 
\subsubsection{Ergodic Sum Capacity}
By summing up (\ref{eq12}), (\ref{eq15}), (\ref{eq22}) and (\ref{eq13}), (\ref{eq16}), (\ref{eq24}), the exact and approximate ESC of the proposed system can be obtained, respectively. 
\subsection{Outage Probability and Diversity Order}
\subsubsection{Outage Probability of U$_1$}
Let R$_1$, R$_3$, R$_d$ are the threshold data rates below which outage occurs for U$_1$, U$_3$, and D$_1$, respectively. Outage will occur in U$_1$ either if it can not decode the information $s_3$ or if it decodes $s_3$ but fails to decode $s_1$. So the OP of U$_1$ can be expressed as \cite{6}
\begin{align}
&\mathcal{P}_{\mathcal{O},1} = 1-\mathcal{P}\left(\log_2\left(1+\gamma_{b1}^{s_{3}}\right)>R_3, \log_2\left(1+\gamma_{b1}^{s_{1}}\right)>R_1\right), \nonumber\\
&= 1-\mathcal{P}\left(\varpi\rho_bg_{b1}>\rho_u\tilde{g}_{21}+1,\frac{\theta_1}{\Lambda_1}\rho_bg_{b1}>\rho_u\tilde{g}_{21}+1\right), \label{eq25} 
\end{align}
where $\varpi=\frac{\theta_3-\theta_1\Lambda_3}{\Lambda_3}$, $\Lambda_1=2^{R_1}-1$, and $\Lambda_3=2^{R_3}-1$. It is noted by analyzing  (\ref{eq25}) that for $\Lambda_3>\frac{\theta_3}{\theta_1}$, the OP becomes $\mathcal{P}_{\mathcal{O},1}=1$ and for $\Lambda_3<\frac{\theta_3}{\theta_1}$, it takes the following form
\begin{align}
\mathcal{P}_{\mathcal{O},1} =& 1-e^{-\frac{1}{\varphi\rho_b\lambda_{b1}}}\left(\frac{\varphi\rho_b\lambda_{b1}}{\varphi\rho_b\lambda_{b1}+\beta}\right), \label{eq26} 
\end{align}  
where $\varphi=\min\left(\frac{\theta_3-\theta_1\Lambda_3}{\Lambda_3}, \frac{\theta_1}{\Lambda_1}\right)$.
\subsubsection{Outage Probability of U$_3$}
If U$_2$ fails to decode $s_3$ or U$_2$ can decode but U$_3$ can not, then outage occurs in U$_3$. Hence, the OP of U$_3$ can be calculated by   
\begin{align}
\mathcal{P}_{\mathcal{O},3} =& 1-\mathcal{P}\left(\log_2\left(1+\gamma_{b2}^{s_{3}}\right)>R_3, \log_2\left(1+\gamma_{23}^{s_{3}}\right)>R_3\right), \nonumber\\
=& 1-\mathcal{P}\left(\varpi\rho_bg_{b2}>\rho_u\tilde{g}_{22}+1,\aleph\rho_ug_{23}>\Lambda_3\right). \label{eq27} 
\end{align}
where $\aleph=\left(\theta_3^\prime-\theta_2\Lambda_3\right)$. From (\ref{eq27}), if $\Lambda_3>\frac{\theta_3}{\theta_1}$ and  $\Lambda_3>\frac{\theta_3^\prime}{\theta_2}$ exist, the OP becomes $\mathcal{P}_{\mathcal{O},3}=1$, whereas for $\Lambda_3<\frac{\theta_3}{\theta_1}$, and $\Lambda_3<\frac{\theta_3^\prime}{\theta_2}$, it can be expressed as
\begin{align}
\mathcal{P}_{\mathcal{O},3} =& 1-e^{-\frac{\Lambda_3}{\aleph\rho_u\lambda_{23}}-\frac{1}{\varpi\rho_b\lambda_{b2}}}\frac{1}{1+\frac{\varsigma_1\lambda_{22}\rho_u}{\varpi\rho_b\lambda_{b2}}}. \label{eq28} 
\end{align} 
\subsubsection{Outage Probability of D$_1$}
The D2D user will be in outage under two conditions; i.e., \textbf{i}. D$_1$ fails to decode U$_3$'s information and \textbf{ii}. D$_1$ decodes $s_3$ but fails to decode $s_2$. So, the OP of D1 can be obtained as
\begin{align}
\mathcal{P}_{\mathcal{O},d} =& 1-\left(\frac{\theta_3^\prime\rho_u g_{2d}}{\theta_2\rho_u g_{2d}+1}>\Lambda_3, \theta_2\rho_u g_{2d}>\Lambda_d\right), \label{eq29} 
\end{align}
where $\Lambda_d=2^{R_d}-1$. For $\Lambda_3>\frac{\theta_3^\prime}{\theta_2}$, the OP becomes $\mathcal{P}_{\mathcal{O},d}=1$. For $\Lambda_3<\frac{\theta_3^\prime}{\theta_2}$, $\mathcal{P}_{\mathcal{O},d}$ is given below
\begin{align}
\mathcal{P}_{\mathcal{O},d} =& 1-e^{-\frac{\Lambda_3}{\aleph\rho_u\lambda_{2d}}-\frac{\Lambda_d}{\theta_2\rho_u\lambda_{2d}}}. \label{eq30} 
\end{align}
\subsubsection{High SNR approximation}
Considering $\rho_b\rightarrow\infty$, and $\rho_u=\varepsilon\rho_b$, the OP of all users can be approximated in high SNR regime where $\varepsilon$ is a relay transmit power controlling variable limiting as $0<\varepsilon\leq1$ \cite{6}. For two cases, i.e., imperfect and perfect SI as well as known interference cancellations (ICs), approximate OPs of U$_1$, U$_3$, and D$_1$ are respectively found as  
\begin{align}
\mathcal{P}_{\mathcal{O},1}^\infty =& \left\{ \begin{array}{rcl}
1-\left(1+\frac{\beta}{\varphi\rho_b\lambda_{b1}}\right)^{-1} & \mbox{for}
& 0<\varsigma_3\leq 1 \\ 1-\left(1+\frac{\varepsilon\varsigma_2\theta_2\lambda_{21}}{\varphi\lambda_{b1}}\right)^{-1} & \mbox{for} & \varsigma_3=0 \end{array}\right. \label{eq31}\\
\mathcal{P}_{\mathcal{O},3}^\infty =& \left\{ \begin{array}{rcl}
1-\left(1+\frac{\varepsilon\varsigma_1\lambda_{22}}{\varpi\lambda_{b2}}\right)^{-1} & \mbox{for}
& 0<\varsigma_1\leq 1 \\ \frac{\Lambda_3}{\aleph\rho_u\lambda_{23}}+\frac{1}{\varpi\rho_b\lambda_{b2}} & \mbox{for} & \varsigma_1=0 \end{array}\right. \label{eq32}
\end{align}
\begin{align}
\mathcal{P}_{\mathcal{O},d}^\infty =& \frac{\Lambda_3}{\aleph\rho_u\lambda_{2d}}+\frac{\Lambda_d}{\theta_2\rho_u\lambda_{2d}}. \label{eq33}
\end{align}
Due to the imperfect IC, OPs of U$_1$ and U$_2$ maintain constant values that create error floors (EFs) in the high SNR. Using $\lim_{\rho_b\rightarrow\infty}\frac{-\log\mathcal{P}_{O}}{\log\rho_b}$ in  (\ref{eq31}), (\ref{eq32}), and (\ref{eq33}), DOs of U$_1$, U$_3$, and D$_1$ under imperfect and perfect ICs are respectively found as 
\begin{figure}[t]
	\centering
	\includegraphics[width=3.72in,height=3in]{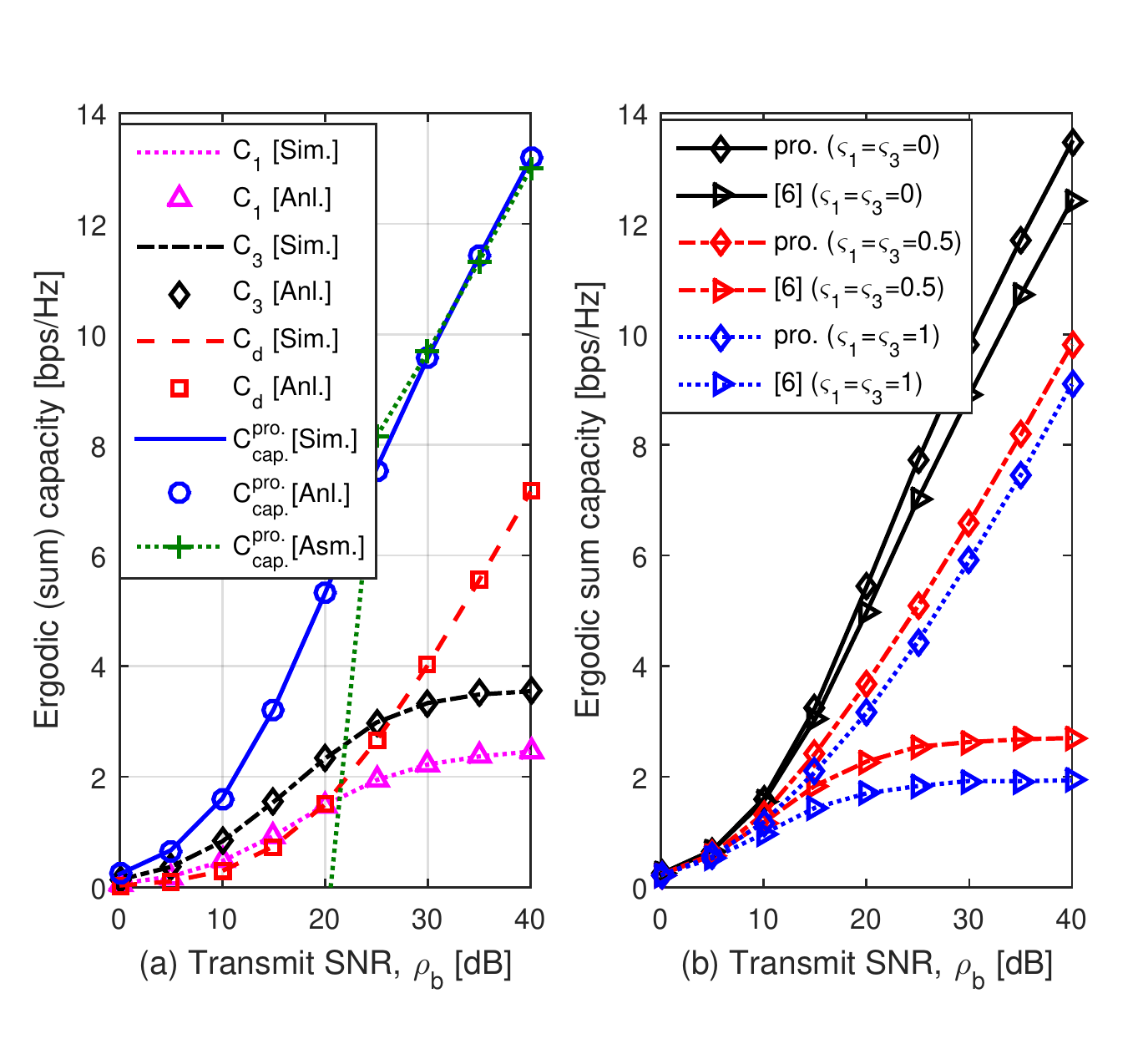}
	\caption{(a) Analytical justification of EC and ESC, (b) ESC comparison between DFC-NOMA and~\cite{6}.}
	\label{Fig2}
\end{figure}
\begin{align}
\mathcal{D}_{\mathcal{O},1}^\infty=0; \mathcal{D}_{\mathcal{O},3}^\infty=0; \mathcal{D}_{\mathcal{O},d}^\infty=1 & \quad \mbox{for}
& 0<\left(\varsigma_1,\varsigma_3\right)\leq 1, \nonumber\\ \mathcal{D}_{\mathcal{O},1}^\infty=0; \mathcal{D}_{\mathcal{O},3}^\infty=1;  \mathcal{D}_{\mathcal{O},d}^\infty=1 & \quad\mbox{for}
&  \varsigma_1=\varsigma_3=0. \label{eq34}
\end{align}
Unknown interference at U$_1$ makes it's DO zero for both perfect and imperfect ICs. Like \cite{6}, U$_3$'s DO is unity under perfect IC and zero under imperfect IC. As there is no interference other than intended NOMA user interference on D$_1$, its DO remains unity under both conditions. 
\section{Numerical Results} 
All findings presented in this section are executed by considering $\rho_b=2\rho_u$, $\lambda_{b1}=\lambda_{2d}=1$, $\lambda_{b2}=\lambda_{21}=\lambda_{23}=0.5$, and $\lambda_{22}=0.3$. In each case, concordance between analytical (Anl.) and simulation (Sim.) plots verifies the accuracy of analyses. EC and ESC performances with respect to (w.r.t) $\rho_b$ are evaluated in Fig.~\ref{Fig2} where $\theta_1=\theta_2=0.05$ and $\theta_3=\theta_3^\prime=0.95$. ECs and ESC of the DFC-NOMA are shown under imperfect IC ($\varsigma_1=0.08^2$, and $\varsigma_3=0.1^2$) in~\ref{Fig2}(a). EC and ESC increase linearly with the increase of $\rho_b$. Due to the large interference effect on SINRs of U$_1$ and U$_3$, their ECs tend to saturate at high $\rho_b$. Contrarily, D$_1$'s EC maintains linear increment as a benefit of performing perfect SIC.  Fig.~\ref{Fig2}(b) displays the performance comparison between DFC-NOMA (pro.) and existing \cite{6} protocols for perfect ($\varsigma_1 = \varsigma_3 = 0$), imperfect ($\varsigma_1 = \varsigma_3 = 0.5$), and no interference ($\varsigma_1 = \varsigma_3 = 1$) cancellations. Proposed protocol outperforms FC-NOMA \cite{6} in terms of ESC for all conditions. 
\par
\begin{figure}[t]
	\centering
	\includegraphics[width=3.72in,height=3in]{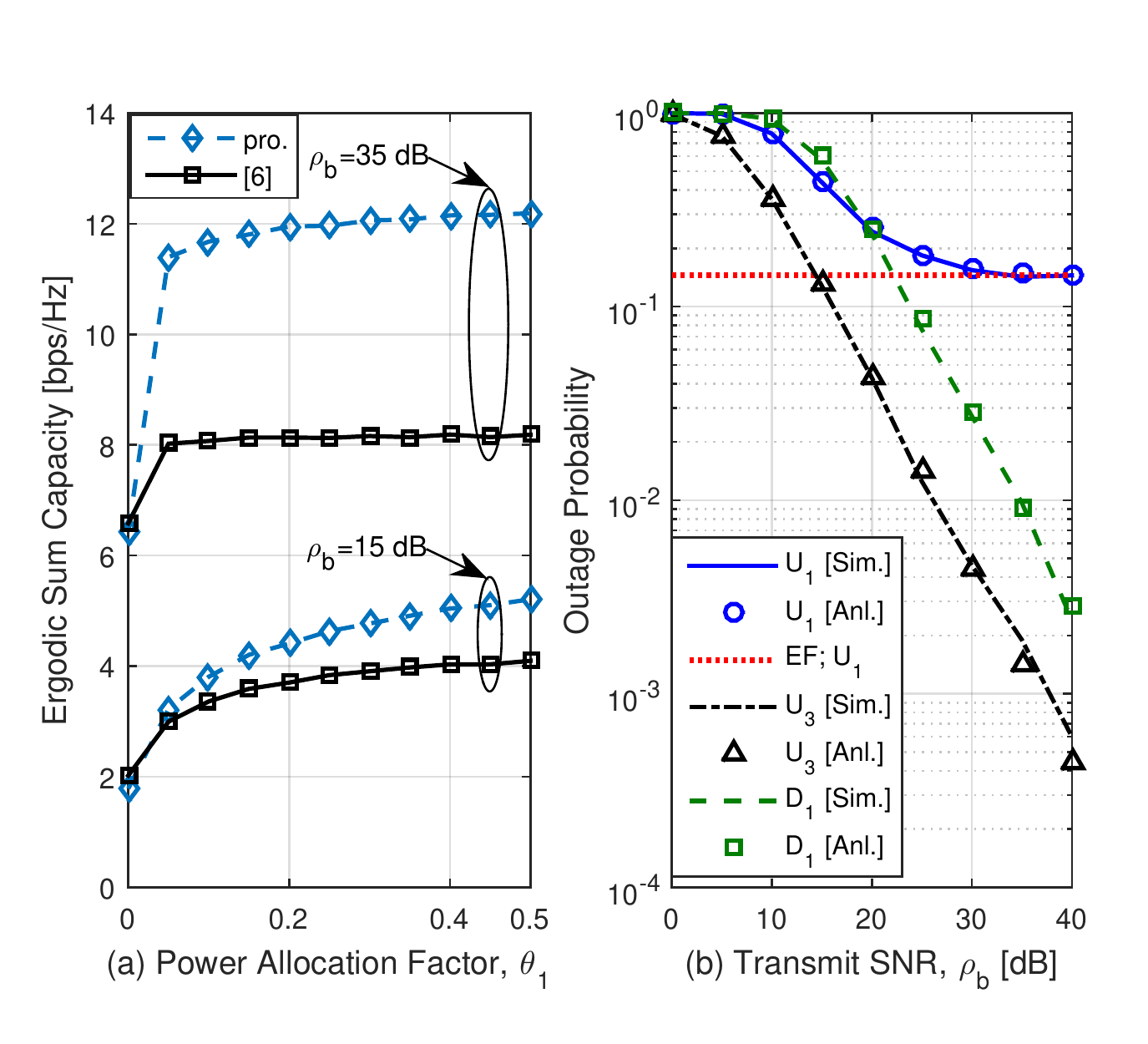}
	\caption{(a) E-SC comparison w.r.t $\theta_1$ between DFC-NOMA and~\cite{6}, (b) OP characteristic of each user with changing $\rho_b$, in DFC-NOMA.}
	\label{Fig3}
\end{figure}

 
Fig.~\ref{Fig3}(a) shows the impact of changing $\theta_1$ on ESC by letting $\theta_1=\theta_2$ and $\theta_3=\theta_3^\prime$ for $\rho_b$=15 and 35 dBs. In all $\theta_1$ and both SNR values, DFC-NOMA poses noticeable ESC gain over \cite{6} under $\varsigma_1=0.08^2$ and $\varsigma_3=0.1^2$. In Fig.~\ref{Fig3}(b), OPs of DFC-NOMA's individual users are presented w.r.t SNR under perfect IC and R$_1$=R$_3$=R$_d$=1.  It is noticed that OPs decrease with the increase of $\rho_b$. OPs of U$_3$ and D$_1$ decrease linearly at high $\rho_b$, whereas U$_1$'s OP saturates due to the presence of interference that creates EF and lead DO to 0.  
\section{Conclusion}
A D2D aided FDCC protocol using NOMA, i.e., DFC-NOMA has been proposed. Considering imperfect IC, the exact and asymptotic expressions of EC, E-SC, OP, and DO of the system have been analyzed over Rayleigh fading channel. By the cost of a slightly increased interference on a near user (U$_1$) than FC-NOMA \cite{6}, DFC-NOMA facilitates simultaneous data transmission to a cellular and a D2D users that helps achieving higher capacity than FC-NOMA. 
\ifCLASSOPTIONcaptionsoff
  \newpage
\fi
\appendices

\bibliographystyle{IEEEtran}
\bibliography{IEEEabrv,D2D_Ref}

\begin{thebibliography}{1}
\providecommand{\url}[1]{#1}
\csname url@samestyle\endcsname
\providecommand{\newblock}{\relax}
\providecommand{\bibinfo}[2]{#2}
\providecommand{\BIBentrySTDinterwordspacing}{\spaceskip=0pt\relax}
\providecommand{\BIBentryALTinterwordstretchfactor}{4}
\providecommand{\BIBentryALTinterwordspacing}{\spaceskip=\fontdimen2\font plus
\BIBentryALTinterwordstretchfactor\fontdimen3\font minus
  \fontdimen4\font\relax}
\providecommand{\BIBforeignlanguage}[2]{{%
\expandafter\ifx\csname l@#1\endcsname\relax
\typeout{** WARNING: IEEEtran.bst: No hyphenation pattern has been}%
\typeout{** loaded for the language `#1'. Using the pattern for}%
\typeout{** the default language instead.}%
\else
\language=\csname l@#1\endcsname
\fi
#2}}
\providecommand{\BIBdecl}{\relax}
\BIBdecl

\bibitem{1}
V.~W. Wong, R.~Schober, D.~W.~K. Ng, and L.-C. Wang, \emph{Key Technologies for
  5G Wireless Systems}.\hskip 1em plus 0.5em minus 0.4em\relax Cambridge
  university press, 2017.

\bibitem{2}
Z.~Ding, H.~Dai, and H.~V. Poor, ``Relay selection for cooperative {NOMA},''
  \emph{IEEE Wireless Commun. Lett.}, vol.~5, no.~4, pp. 416--419, Aug. 2016.

\bibitem{3}
M.~F. Kader, S.~Y. Shin, and V.~C.~M. Leung, ``Full-duplex non-orthogonal
  multiple access in cooperative relay sharing for {5G} systems,'' \emph{IEEE
  Trans. Veh. Technol.}, vol.~67, no.~7, pp. 5831--5840, Jul. 2018.

\bibitem{4}
Z.~Zhang, Z.~Ma, M.~Xiao, Z.~Ding, and P.~Fan, ``Full-duplex
  device-to-device-aided cooperative nonorthogonal multiple access,''
  \emph{IEEE Trans. Veh. Technol.}, vol.~66, no.~5, pp. 4467--4471, May 2017.

\bibitem{5}
N.~Madani and S.~Sodagari, ``Performance analysis of non-orthogonal multiple
  access with underlaid device-to-device communications,'' \emph{IEEE Access},
  vol.~6, pp. 39\,820--39\,826, 2018.

\bibitem{6}
C.~Zhong and Z.~Zhang, ``Non-orthogonal multiple access with cooperative
  full-duplex relaying,'' \emph{IEEE Commun. Lett.}, vol.~20, no.~12, pp.
  2478--2481, Dec. 2016.

\bibitem{7}
T.~Chen and S.~Liu, ``A multi-stage self-interference canceller for full-duplex
  wireless communications,'' in \emph{2015 IEEE Global Communications
  Conference (GLOBECOM)}, Dec. 2015, pp. 1--6.

\bibitem{8}
I.~S. Gradshteyn and I.~M. Ryzhik, \emph{Table of integrals, series and
  products}.\hskip 1em plus 0.5em minus 0.4em\relax 7th edn. New York, NY, USA:
  Academic, 2007.

\end{thebibliography}
\end{document}